\documentclass[10pt,twocolumn,letterpaper]{article}

\usepackage{cvpr}
\usepackage{times}
\usepackage{epsfig}
\usepackage{graphicx}
\usepackage{amsmath}
\usepackage{amssymb}
\usepackage[pagebackref=true,breaklinks=true,letterpaper=true,colorlinks,bookmarks=false]{hyperref}
\usepackage{caption} 
\def\ie{\emph{i.e.}}

\newcommand{\z}{\mathbf{z}}
\newcommand{\x}{\mathbf{x}}

\newcommand{\E}{\mathbb{E}}

\newcommand{\enc}{\mathbf{E}_\phi}
\newcommand{\dec}{\mathbf{D}_\psi}
\newcommand{\dis}{\mathbf{C}_\omega}
\newcommand{\quantize}{\mathbf{Q}}
\newcommand{\entropy}{\mathbf{H}}
\newcommand{\distortion}{\mathbf{d}}
\newcommand{\vgg}{\mathbf{\sigma}}

\newcommand\blfootnote[1]{%
  \begingroup
  \renewcommand\thefootnote{}\footnote{#1}%
  \addtocounter{footnote}{-1}%
  \endgroup
}

\cvprfinalcopy 

\def\cvprPaperID{24} 

\ifcvprfinal\fi
\begin{document}

\title{Adversarial Distortion for Learned Video Compression}

\author{Vijay Veerabadran\textsuperscript{$\dagger$3}, Reza Pourreza\textsuperscript{1}, Amirhossein Habibian\textsuperscript{2}, Taco Cohen\textsuperscript{2}\\
\textsuperscript{1} Qualcomm AI Research, Qualcomm Technologies, Inc. \\
\textsuperscript{2} Qualcomm AI Research, Qualcomm Technologies Netherlands B.V.\\
\textsuperscript{3} Department of Cognitive Science, University of California San Diego\\
}

\maketitle
\thispagestyle{empty}

\begin{abstract}
In this paper, we present a novel adversarial lossy video compression model. At extremely low bit-rates, standard video coding schemes suffer from unpleasant reconstruction artifacts such as blocking, ringing etc. Existing learned neural approaches to video compression have achieved reasonable success on reducing the bit-rate for efficient transmission
and reduce the impact of artifacts to an extent. However, they still tend to produce blurred results under extreme compression.
In this paper, we present a deep adversarial learned video compression model that minimizes an auxiliary adversarial distortion objective. We find this adversarial objective to correlate better with human perceptual quality judgement relative to traditional quality metrics such as MS-SSIM and PSNR.
Our experiments using a state-of-the-art learned video compression system demonstrate a reduction of perceptual artifacts and reconstruction of detail lost especially under extremely high compression.
\end{abstract}
\vspace{-.3cm}

\blfootnote{$^\dagger$ Work completed during internship at Qualcomm Technologies, Inc.\\Qualcomm AI Research is an initiative of Qualcomm Technologies, Inc.}
\vspace{-1cm}

\section{Introduction}
\label{sec:introduction}
As the resolution of digitally recorded and streamed videos keeps growing, there is an increasing demand for video compression algorithms that enable fast transmission of videos without loss in Quality-of-Experience.
While current video codecs can encode video at low bitrates, this usually results in unpleasant compression artifacts \cite{zeng2014characterizing, pourreza2019recognizing}.
The application of deep neural networks to develop learned video compression algorithms as explored in recent art \cite{tschannen2018deep, rippel2019learned, lu2019dvc, habibian2019video,chen2019learning, chen2017deepcoder, djelouah2019neural} produces promising results at solving this issue of perceptual artifacts.
However, due to the use of distortion metrics such as MS-SSIM~\cite{msssim} and MSE, the reconstructions tend to be blurry \cite{blau2018perception}.
%
\begin{figure}[h]
    \includegraphics[width=\linewidth]{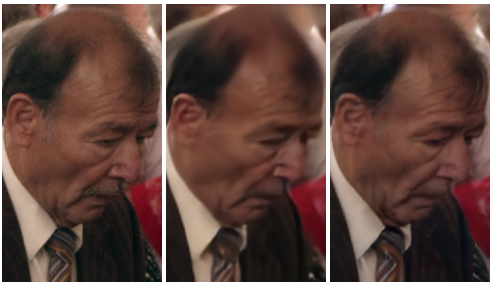}
    \vspace{-0.2cm}
    \begin{tabular}{p{.28\linewidth}p{.28\linewidth}p{.28\linewidth}}
        \centering (a) & \centering (b) & \centering (c)
    \end{tabular}
    \caption{Demonstration of the effectiveness of training with adversarial loss. (a) uncompressed frame, learned compression~\cite{habibian2019video} via (b) MS-SSIM distortion, (c) adversarial distortion, at similar bitrates.
    \tiny{[See Fig.~\ref{fig:screenshots} cation for license information]}}
    \label{fig:intro_screenshot}
    \vspace{-0.5cm}
\end{figure}
Generative Adversarial Networks (GANs) have been shown to be capable of producing highly realistic images and videos from random noise inputs \cite{karras2019style,karras2017progressive,zhang2018self,brock2018large,clark2019efficient}.
This suggests that the GAN objective more accurately reflects image/video quality as perceived by humans.
Indeed the work of \cite{agustsson2019generative} has shown that GANs can be used for low-rate high-quality image compression, by augmenting the rate/distortion loss with an adversarial loss.
However, so far there is little work on the application of adversarial losses to video compression due to scaling issues. 

We tackle the scaling issue via factorization of our adversarial discriminator into smaller neural network components and show results that demonstrate our compression system's relatively improved perceptual quality even under extreme compression (see Fig.~\ref{fig:intro_screenshot}).
Our model is based on the one proposed by \cite{habibian2019video}, which is a 3D autoencoder with discrete latents and a PixelCNN++\cite{van2016conditional} prior, trained end-to-end to optimize a rate/distortion loss. Our contributions shall be applicable to other learned video compression systems in general.
We also present an ablation study resulting from our search across various formulations of GANs in terms of their architecture and loss functions within the context of lossy video compression.
The contributions of this paper are: \emph{i)} we propose adversarial loss to improve the perceptual quality of learned video compression, \emph{ii)} we study techniques to improve the training stability using adversarial loss, \emph{iii)} we study a spatial-temporal factorization of discriminator to enable end-to-end training of deep video compression networks. 

\section{Learned Video Compression}
\label{sec:background}
\begin{figure}[t!]
    \includegraphics[width=.9\linewidth]{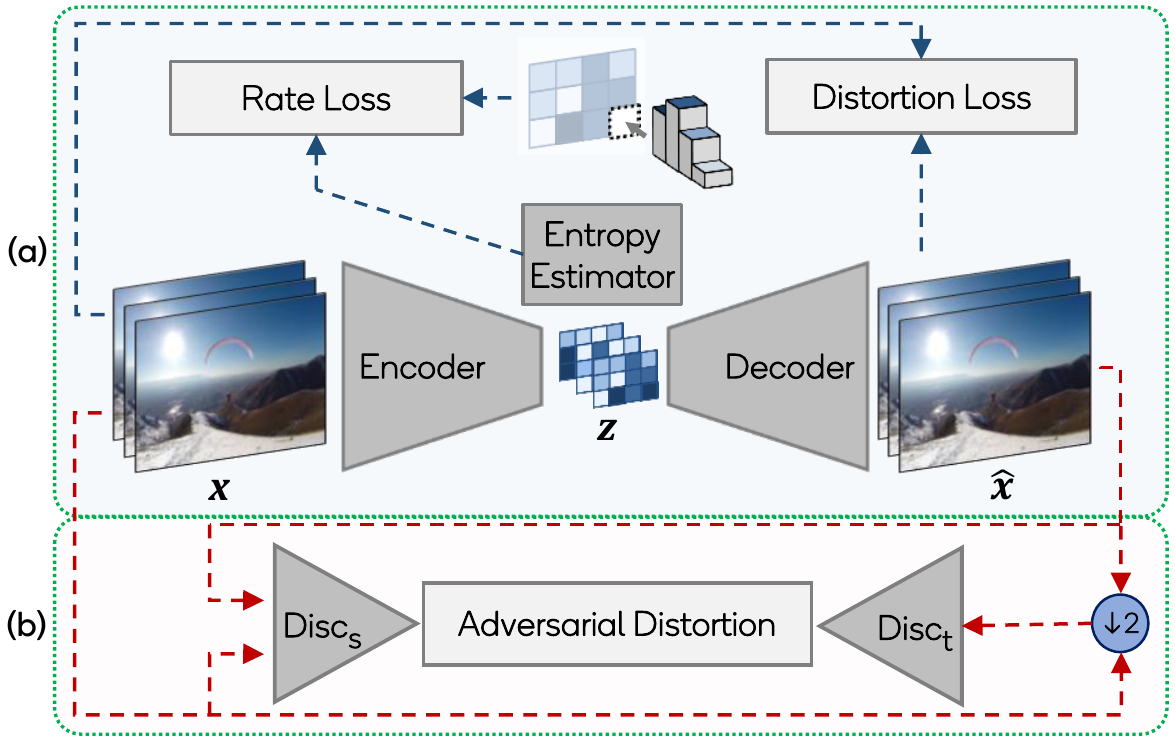}
    \caption{Lossy video compression with adversarial distortion, (a) learned video compression component, (b) adversarial distortion components, where Disc\textsubscript{s} and Disc\textsubscript{t} represent the spatial and the spatio-temporal discriminators.
    }
    \tiny{[Frames by Ambrose Productions \texttt{CC BY-SA 3.0} \url{https://creativecommons.org/licenses/by/3.0/legalcode}, via YouTube]}
    \label{fig:net4_gan_architecture}
    \vspace{-0.5cm}
\end{figure}
A learned video compression typically consists of an encoder, decoder, entropy estimator and distortion estimator. All the components are trained end-to-end on a collection of videos $\x$.

Encoder maps an input sequence $\x$ into a latent representation $\z = \enc (\x)$. Encoder is a stack of convolutional layers with several down-samplings
that reduce the input dimension. In the last layer, encoder employs a quantisation function on the activations to reduce the bit-width of latents $\z = \quantize(\tilde{\z})$. Quantizer maps each element, or group of elements, in activations $\tilde{\z}_i$ to a discrete symbol $\z_i \in \{0, \dots, L\}$. Learning discrete representation, as a non-differentiable function, requires adding uniform noise or soft assignment as an approximation. The decoder, which is a stack of convolutional layers with several up-samplings, reconstructs the video given discrete latents $\hat{\x} = \dec (\z)$. 

Entropy estimator predicts the average number of bits needed to encode latents using a lossless entropy coding schema such as Huffman or arithmetic coding. The bit rate is measured as the cross-entropy between the true distribution of latents $p(\z)$ and a density estimated by $P_{\theta}(\z)$ as in Eq.~\ref{eq:entropy}. The density estimator $P_{\theta}(\z)$ is parameterized as a neural network usually with an auto-regressive architecture~\ie\ PixelCNN++~\cite{van2016conditional}.
\begin{equation}
    \label{eq:entropy}
    \entropy(\z) = \E_{\z \sim p(\z)}[-\log(P_{\theta}(\z))]
\end{equation}
Distortion loss measures the difference between the input and reconstructed videos $\distortion(\x, \hat{\x})$ usually by pixel-wise distances, such as $\ell_1$ and $\ell_2$, or by more sophisticated metrics such as MS-SSIM. All the aforementioned components are trained end-to-end by minimizing a rate-distortion trade-off as the loss function:
\begin{equation}
    \label{eq:loss}
    \distortion(\x, \dec(\enc(\x)) + \beta \entropy(\z)
\end{equation}

\section{Adversarial Distortion Loss}
\label{sec:method}
The distortion, measured by pixel-wise metrics~\ie\ $\ell_1$, $\ell_2$, and MS-SSIM, are often not perfectly aligned with perceptual quality. Recent work~\cite{blau2018perception} mathematically proves that the distortion and perceptual quality are at odds with each other and minimizing the mean distortion leads to a decrease in perceptual quality. Instead of solely relying on pixel-wise distances, we define the distortion as an adversarial loss between the decoder and a discriminator $\dis$. This setting can be interpreted as training a conditional GAN, where the decoder $\dec$ learns to generate a video given the encoded latents $\z$. The discriminator encourages the decoder to generate videos which reside on the data manifold that improves perceptual quality.

\begin{table}[t!]
  \resizebox{\linewidth}{!}{
  \centering
  \begin{tabular}{|l|c|c|c|}
    \hline
    & $f$ & $g$ & $h$\\
    \hline
    Minimax \cite{goodfellow2014generative} & $-\log(1 + e^{-y})$ & $-y - \log(1 + e^{-y})$ & $-y - \log(1 + e^{-y})$\\
    Wasserstein \cite{arjovsky2017wasserstein} & $y$ & $-y$ & $-y$\\
    Least Squares \cite{mao2017least} & $-(y - 1)^2$ & $-y^2$ & $(y - 1)^2$\\
    \hline
  \end{tabular}
  }
  \caption{Component functions for a few adversarial losses.
  }
  \vspace{-0.5cm}
  \label{tab:loss_functions}
\end{table}

\subsection{Stable adversarial training}
\textbf{Adversarial loss}
Adversarial loss 
can be defined in various formulations depending on how to specify the component functions $f$, $g$ and $h$:
%
\begin{align}
    \label{eqn:gan_loss}
    \max_{\dis}\;&\E_{\x}[f(\dis(\x))] + \E_{\hat\x}[g(\dis(\hat\x))]\\
    \min_{\dec}\;&\E_{\hat\x}[h(\dis(\hat\x))]
\end{align}
Table~\ref{tab:loss_functions} specifies the component functions for several widely used GAN formulations. 
We investigate the impact of each formulation on training stability as they have different loss landscapes and gradient behavior; finding the best formulation is hence non-trivial. 
Minimax loss \cite{goodfellow2014generative} and Wasserstein loss \cite{arjovsky2017wasserstein} resulted in fairly decent improvements in terms of reconstruction quality, but we noticed the training to be unstable and time consuming. 
We also experimented with the Least Squares \cite{mao2017least} and Relativistic~\cite{jolicoeur2018relativistic} formulations.
Both of these formulations resulted in stable adversarial training. Among these two choices, our best formulation was the Least Squares loss that generated higher quality videos (see section \ref{sec:ablation}).

\textbf{Perceptual loss}
As a way of further stabilizing our model's adversarial training, we incorporated a semantic loss \cite{ledig2017photo,wang2018esrgan} that minimizes the $\ell_1$ of the difference between framewise VGG-19 representations of $\x$ and $\hat\x$.
This semantic loss resulted in faster and more stable training of our adversarial video compression model.

\subsection{Factorized spatial-temporal discriminators}
Recent work in training GANs to generate videos points out the advantage of scaling up the training, \ie, using larger batch sizes, deeper models etc. \cite{clark2019efficient}. However, due to scalability issues relating to working with video data and the models size, we faced difficulty in jointly training all components. In this case, our two choices to scale up our training were: \emph{(i)} finetune the decoder using an adversarial distortion and fixing the prior and encoder, hence loading only the adversarial distortion components in memory, and \emph{(ii)} factorizing our model into smaller 
components that enable large-scale training. While analyzing the compression performance of the above two choices,
we observed that the latter produced higher quality reconstruction at the same bit-rates. In order to scale up joint adversarial training for our complete model, we resorted to factorizing the discriminator into two smaller spatial and spatio-temporal discriminators. Both these discriminators were formulated as LSGAN discriminators. The average loss from these discriminators was used to train the decoder. 

Putting all together, we train our model end-to-end by optimizing the rate-distortion trade-off Eq.~\ref{eq:loss} using the following distortion loss:
\begin{equation}
    \label{eq:total_loss}
    \distortion(\x, \hat\x) = \alpha \left \| \x - \hat{\x} \right \|_2 + \gamma \left \| \vgg(\x) - \vgg(\hat{\x}) \right \|_1 + \rho [(\dis(\hat{\x})-1)^2]
\end{equation}
%

\noindent where $\vgg$ represents the VGG-19 features\footnote{We used features from the $4^{th}$ convolution  before  the  5$^{th}$ max-pooling layer of an ImageNet-trained VGG-19 network.}. These design choices are summarized in our architecture in Fig. \ref{fig:net4_gan_architecture}.

\section{Experiments}
\label{sec:experiments}
\begin{figure}[t!]
    \centering
    \includegraphics[width=.88\linewidth]{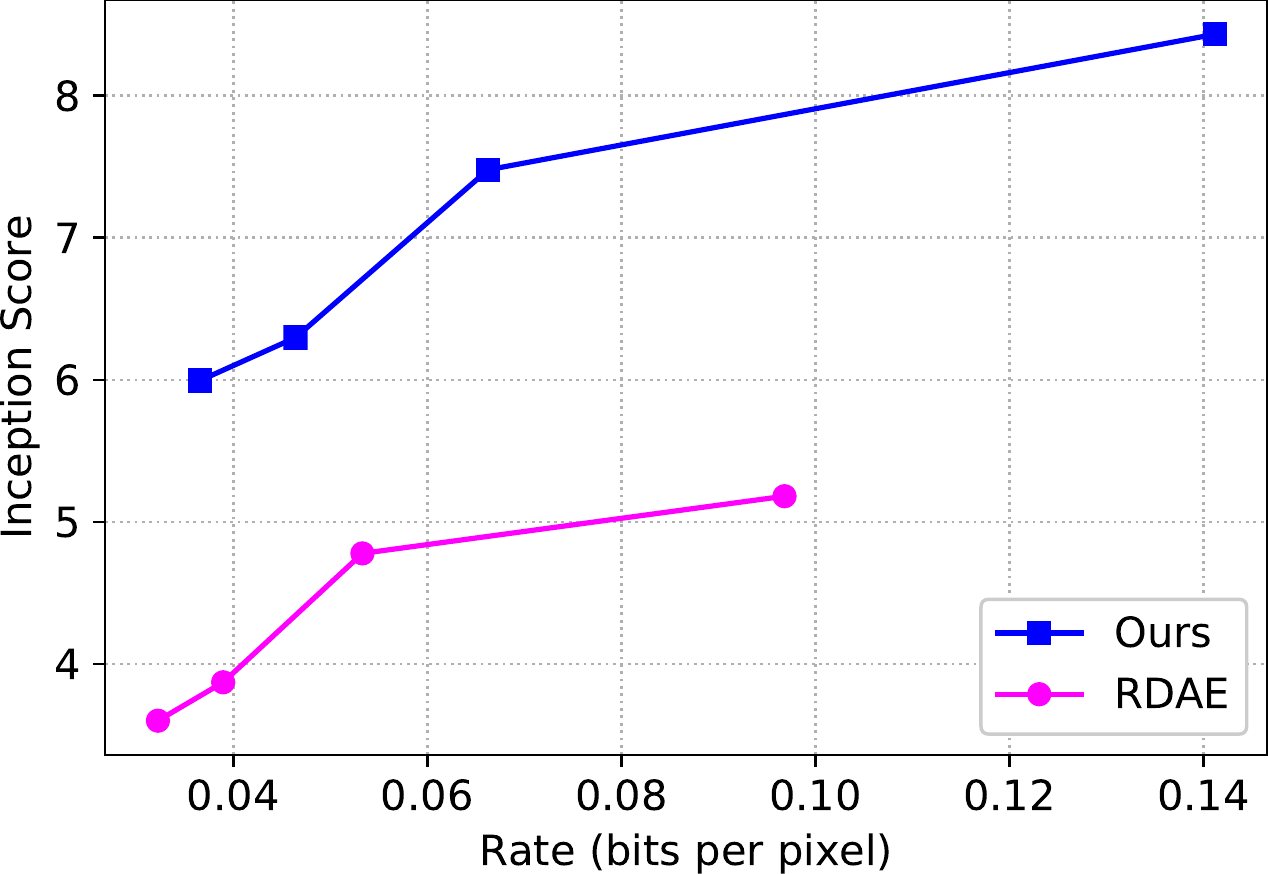}
    \caption{Comparison of RDAE and our method on Kinetics validation set.}
    \label{fig:inception}
    \vspace{-0.4cm}
\end{figure}

\subsection{Experimental setup}
\textbf{Network architecture}
We demonstrate the impact of our adversarial training on the Rate-Distortion Autoencoder \cite{habibian2019video} (RDAE) model which achieves state-of-the-art video compression performance compared to other learned \cite{lu2019dvc,wu2018video} 
compression methods in terms of
MS-SSIM at various bit-rates. 
Further details on the architecture of RDAE's encoder, decoder, entropy estimator and quantizer can be found in \cite{habibian2019video}.
We employed a 2D ResNet-34~\cite{He2015} (trained on ImageNet) and a 3D ResNet~\cite{hara3dcnns} (trained from scratch) as our spatial and spatio-temporal discriminators respectively.
We spatially downsize our spatio-temporal discriminator's input by half
in order to save memory, however, we did not temporally subsample our spatial discriminator's input.

\textbf{Dataset}
We created a dataset sourced from Kinetics400~\cite{Kinetics400} by selecting the first 16 frames from each high-quality video and downsampling them to alleviate the existing compression artifacts resulting in a total of 93750 videos for training and 5687 videos for validation. We used random $160\times160$ crops for training and used full-size frames for validation.
We used UVG \cite{UVG} as the test dataset for comparisons with other methods.

\begin{figure}[t!]
    \centering
    \includegraphics[width=.95\linewidth]{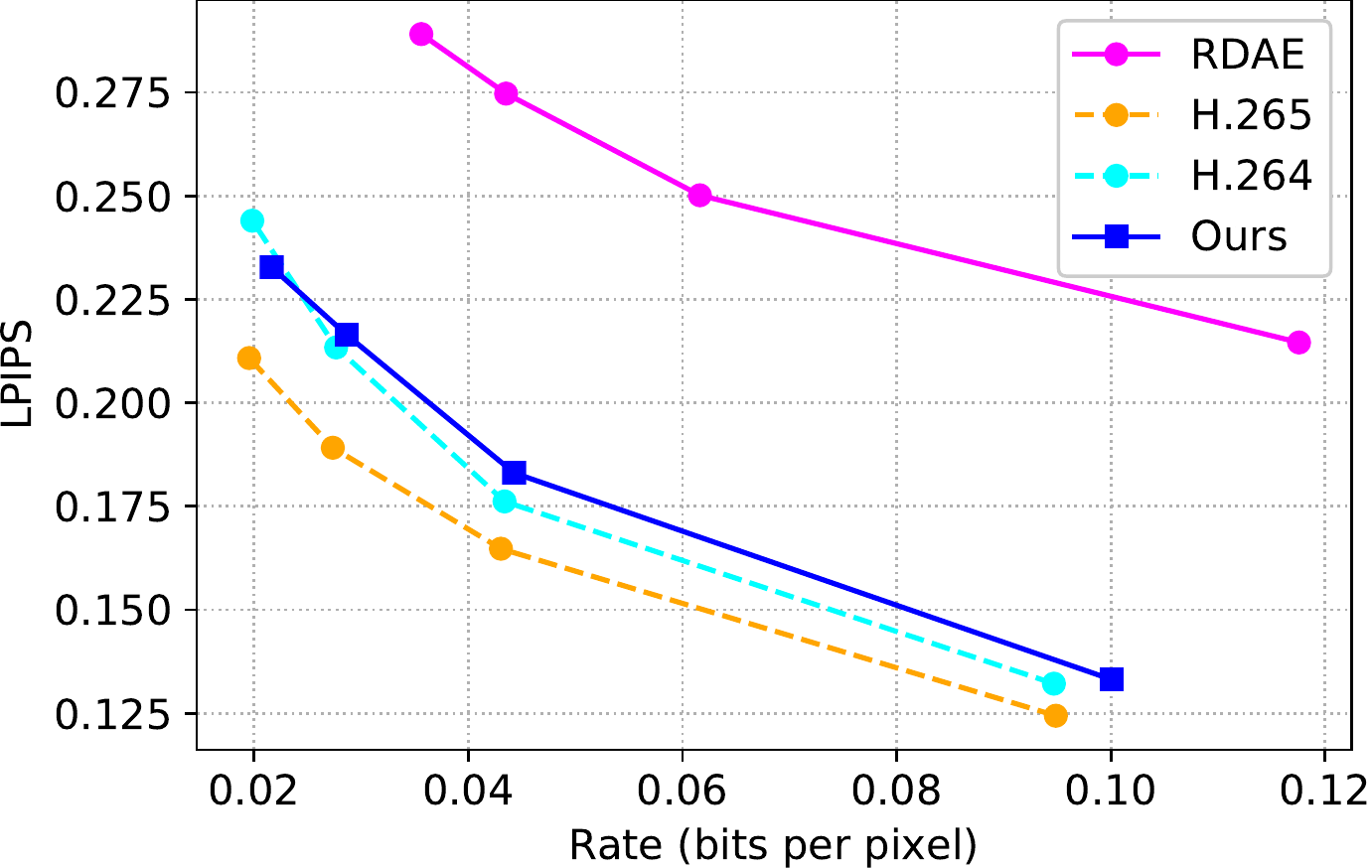}
    \caption{LPIPS comparisons of H.264, H.265, RDAE, and our method on UVG.}
    \label{fig:lpips}
    \vspace{-0.4cm}
\end{figure}

\textbf{Implementation details}
We pretrained our video compression model using rate-loss and MSE distortion loss to speed up adversarial training; this step provided a good initialization for adversarially training decoder weights. 
Our hyperparameter choices for optimizing Eq.~\ref{eq:total_loss} are: $\alpha=0.005$, $\gamma=0.1$, and $\rho =0.0001$.
We used a batch size of 37, for a total of 12 epochs. We trained our network with 4 values of $\beta=\{0.1, 0.3, 0.5, 0.7\}$ to obtain a rate-distortion curve.
All models were trained using Adam optimizer~\cite{kingmaAdamMethodStochastic2015} 
with learning rate of $10^{-4}$, $\beta_1=0.9$ and $\beta_2=0.999$. 

\begin{figure*}
\begin{center}
\setlength{\tabcolsep}{1pt} 
\renewcommand{\arraystretch}{.5} 
  \begin{tabular}{cc}
    \includegraphics[width=.49\textwidth]{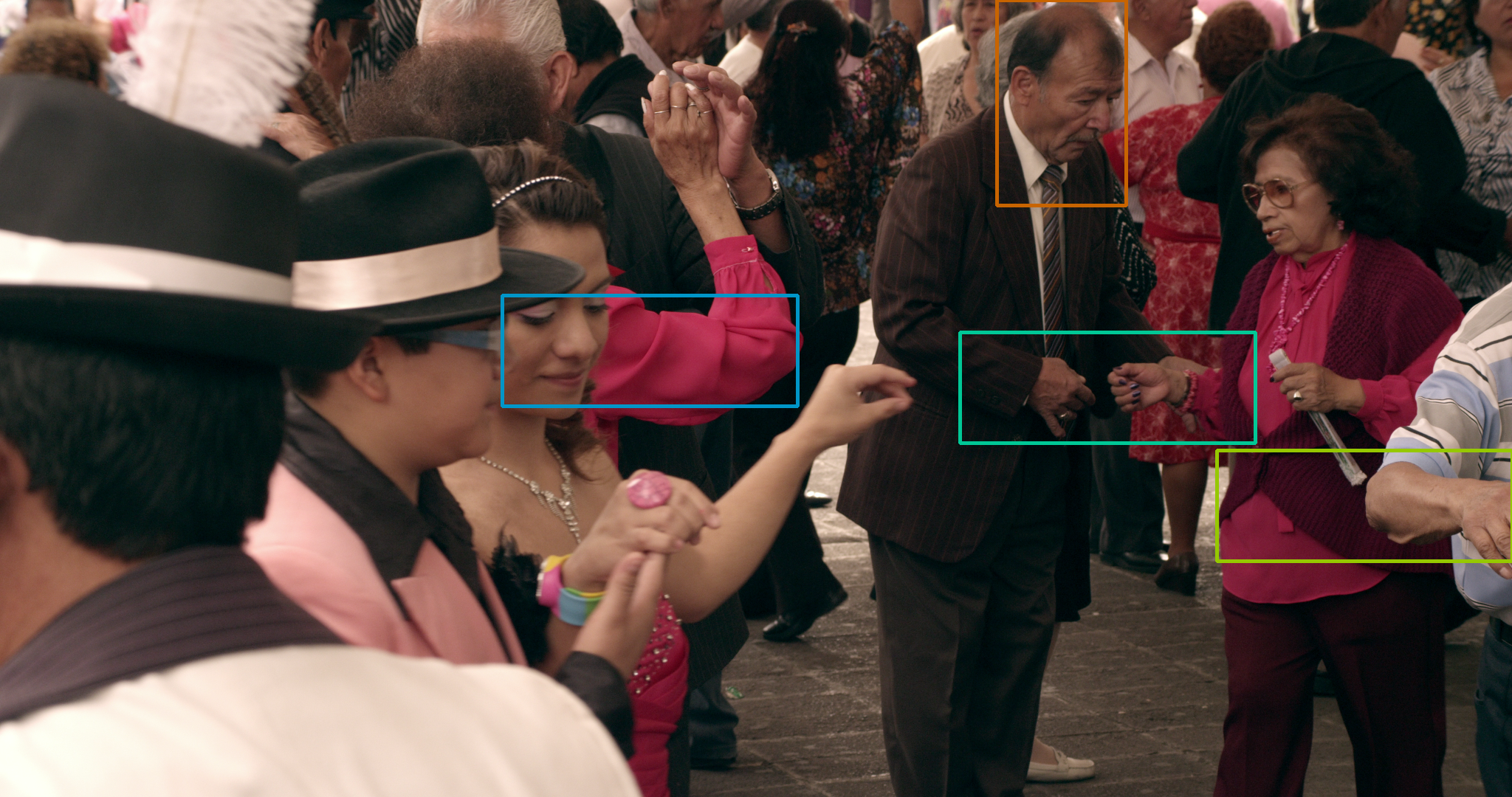} & 
    \includegraphics[width=.485\textwidth]{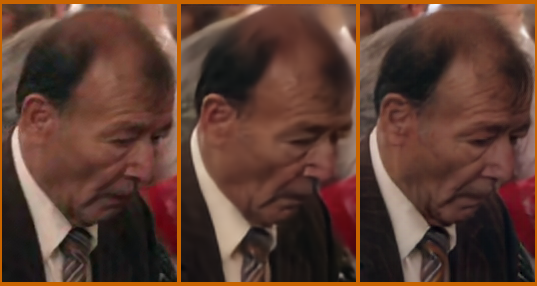} \\
    (a) Uncompressed frame & \\
  \end{tabular}
  \begin{tabular}{c}
    \includegraphics[width=.99\textwidth]{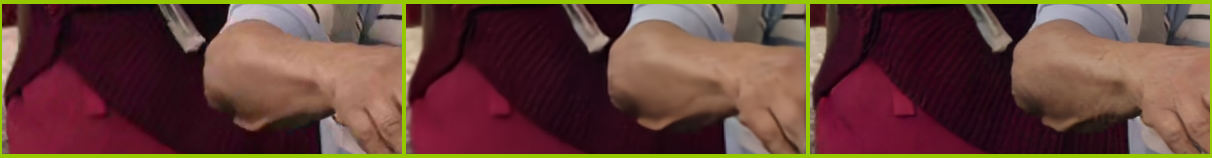} \\
    \includegraphics[width=.99\textwidth]{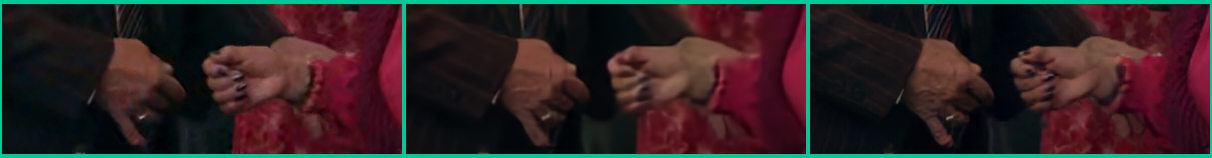} \\
    \includegraphics[width=.99\textwidth]{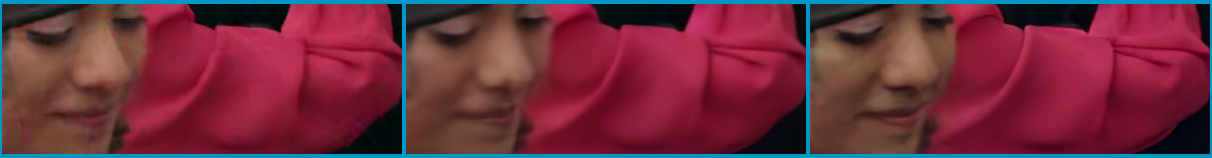} \\
  \end{tabular}
  \begin{tabular}{p{.33\textwidth}p{.33\textwidth}p{.33\textwidth}}
    \centering (b) H.265, 0.0294 bpp & \centering (c) RDAE, 0.0339 bpp & \centering (d) Ours, 0.0309 bpp
  \end{tabular}
\end{center}
\vspace{-0.4cm}
\caption{Visual comparison of H.265, RDAE, and our method at a  comparable  bitrate. (a) shows an uncompressed frame (frame 49 of Netflix Tango in Netflix El Fuente~\cite{xiph}), (b), (c), and (d) show the close-ups of the reconstructed frame from H.265, RDAE, and our method, respectively. Our method is void of the compression artifacts present in H.265 and oversmoothing present in RDAE under low bit-rates.}
\tiny[Frame 49 of Tango video from Netflix Tango in Netflix El Fuente, produced by Netflix, with \texttt{CC BY-NC-ND 4.0} license: {\texttt{https://media.xiph.org/video/derf/ElFuente/Netflix\_Tango\_Copyright.txt}}]
\label{fig:screenshots}
\vspace{-0.4cm}
\end{figure*}

\subsection{Results}

\textbf{Comparison to state of the art}
In this section, we compare the performance of our method 
with a learned video compression method RDAE~\cite{habibian2019video} as well as two non-learned codecs H.265~\cite{hevc} and H.264~\cite{avc}. 
Fig.~\ref{fig:inception} shows the comparison of our method with RDAE in terms of Inception Score~\cite{inceptionscore} (IS, $\uparrow$)
on Kinetics validation set. Incorporation of adversarial training improves IS, and hence the perceptual salience of decoded videos by a great extent. 

Due to the small number of videos present in UVG, we are unable to accurately compute IS on this dataset; we report a framewise Learned Perceptual Image Patch Similarity~\cite{lpips} (LPIPS, $\downarrow$) score on UVG for this reason.
Fig.~\ref{fig:lpips} shows the comparison of our method with RDAE, H.264 and H.265
in terms of LPIPS on UVG dataset. 
We used \texttt{ffmpeg}~\cite{ffmpeg} implementation of H.265 and H.264.
From this analysis, we observe that adversarial training improves the perceptual quality of RDAE, resulting in a lower LPIPS score. 
Our model, albeit underperforms H.265, closes the gap between the learned methods and H.265 in terms of LPIPS to a high extent. 
In Fig.~\ref{fig:screenshots} we compare the visual quality of our lowest rate model with H.265 and RDAE at a similar rate. 
We can see that our result is relatively free from the compression artifacts usually present in H.265 and RDAE at low bitrate. 

\textbf{Ablation study} \label{sec:ablation}
We trained video compression models separately on each of the 4 GAN loss formulations mentioned in section~\ref{sec:method} along with a stabilizing pixel-wise 
$\ell_1$- or 
$\ell_2$-norm of the distance between $\x$ and $\hat\x$ (8 GAN models in total, WGAN not reported due to unstable training).
We report a summary of these experiments in
Table~\ref{tab:ablation_table} and we decided to use the best performing LSGAN with $\ell_2$ pixel-wise loss for our video compression model.

\begin{table}[h]
    \small
    \centering
    \begin{tabular}{|c|c|c|c|}
        \hline
        \textbf{GAN Type} & \textbf{Pixel Loss} & \textbf{MS-SSIM} & \textbf{PSNR (dB)} \\
        \hline
        DCGAN \cite{goodfellow2014generative} & L1 & 0.957 & 26.655 \\
        DCGAN & L2 & 0.958 & 26.862 \\
        RaGAN \cite{jolicoeur2018relativistic}& L1  & 0.957 & 26.554 \\
        RaGAN & L2 & 0.957 & 26.625 \\
        LSGAN \cite{mao2017least}& L1 & 0.96 & 26.905 \\
        LSGAN & L2 & \textbf{0.961} & \textbf{27.032} \\
        \hline
    \end{tabular}
    \caption{Ablation study on different GAN losses, PSNR and MS-SSIM Comparisons on Kinetics validation set.}
    \label{tab:ablation_table}
    \vspace{-.6cm}
\end{table}






\section{Conclusion}
In this paper, we have presented a new deep adversarial lossy video compression algorithm that outperforms state-of-the-art learned video compression systems in terms of visual quality. By employing adversarial training for the decoder, we demonstrate reduction in the perceptual artifacts (especially under very low bit-rates) typically present in the reconstructed output. We have also presented an ablation study of our design choices that resulted in our final adversarial compression model. 
{\small
\bibliographystyle{ieee_fullname}
\bibliography{refs}
}

\end{document}